\documentstyle[preprint,eqsecnum,aps]{revtex}
\input epsf
\newcommand{\extraspace}{\addtolength{\abovedisplayskip}{2mm}
                         \addtolength{\belowdisplayskip}{2mm}
                         \addtolength{\abovedisplayshortskip}{2mm}
                         \addtolength{\belowdisplayshortskip}{2mm}}

\newcommand{\be}{\begin{equation}\extraspace}
\newcommand{\ee}{\end{equation}}
\newcommand{\bea}{\begin{eqnarray}\extraspace}
\newcommand{\eea}{\end{eqnarray}}
\newcommand{\nonu}{\nonumber \\[2mm]}

\newcommand{\ipi}{i {\pi \over 2}}
\newcommand{\th}{\theta}
\newcommand{\qh}{{\theta \over 2}}

\newcommand{\STR}{\rule[-5.5mm]{0mm}{13mm}}

\begin{document}

\title{\large
Reflection Matrices for Integrable $N=1$ Supersymmetric Theories} 
\author{ M. Moriconi {$^\diamondsuit$}\footnotetext
{{$\diamondsuit \;\;$\small e.mail: marco@puhep1.princeton.edu}}} 
\address{ Physics Department, Princeton University
\\Jadwin Hall, Princeton, NJ 08544, U.S.A.}

\author{K. Schoutens {$^\heartsuit$} \footnotetext
{{$\heartsuit \;\;$\small e.mail: kjschout@phys.uva.nl}}}
\address{ Institute for Theoretical Physics, University of Amsterdam\\
Valckenierstraat 65, 1018 XE Amsterdam, The Netherlands}

\maketitle
\begin{abstract}
We study two-dimensional integrable $N=1$ supersymmetric 
theories (without topological charges) in the presence of 
a boundary. We find a universal ratio between the reflection
amplitudes for particles that are related by supersymmetry
and we propose exact reflection matrices for 
the supersymmetric extensions of the multi-component
Yang-Lee models and for 
the breather multiplets of the supersymmetric sine-Gordon theory.
We point out the connection between our reflection matrices
and the classical boundary actions for the supersymmetric
sine-Gordon theory as constructed by Inami, Odake and Zhang
\cite{IOZ}.

\end{abstract}

\vfill

\noindent PUPT-1619

\noindent ITFA-96-09

\newpage

\section{Introduction}

Quantum field theory (QFT) has been very successful in the 
description of physical phenomena in a wide area, ranging from particle
physics at the one end to statistical mechanics and condensed matter
physics at the other. When working out applications of QFT
to physical systems, one typically finds that, while realistic
theories tend to be hard to analyze, the ones that we can
solve in closed form tend to be not so realistic. However, a number
of successes (in particular some recent ones, where exactly solvable 
QFT's were applied to some two dimensional condensed matter systems) have 
clearly shown that the situation is not so bad after all, and that 
there is a non-empty overlap between `realistic' and `solvable' QFT's.
Especially promising in this respect are the integrable QFT's in two 
dimensions. Due to the existence of an infinite number of commuting 
conserved charges, these are highly constrained theories that display 
some very nice features, in particular the absence of particle production 
in scattering processes and the factorizability of the $S$-matrix.

It is interesting to consider QFT's which, apart from being integrable, 
are at the same time supersymmetric. In many examples \cite{Za1,FMV},
the $N=1$ or $N=2$ supersymmetry algebra contains topological charges, 
and the fundamental particles should be viewed as kinks in a non-trivial 
potential. It is possible, however, to have $N=1$ supersymmetric 
theories without topological charges \cite{SW,Sch,Ahn1}, 
and it is on this case that
we shall focus in this paper. One motivation for including
supersymmetry in integrable theories comes from particle 
physics and the superstring connection. In addition, it may
be observed that supersymmetric factorizable $S$-matrices are 
among the simplest ones that are non-diagonal. Certain procedures that 
are problematic for non-diagonal $S$-matrices, such as 
for example the Thermodynamic Bethe Ansatz (TBA), become manageable
when the non-diagonal $S$-matrices are controled by an underlying
symmetry such as supersymmetry \cite{FI,Ahn2,MS}. This point was made 
very clearly in \cite{KS,Ahn2,MS}, where it was shown that $N=1$ 
supersymmetry leads to a `free fermion condition' for certain two-particle 
$S$-matrices, and that this free fermion condition makes it possible 
to perform the TBA analysis in closed form.

In \cite{SW,Sch,Ahn1} the exact $S$-matrices were found for the $N=1$ 
supersymmetric Yang-Lee model and its multicomponent generalizations, 
and for the bound state multiplets of the $N=1$ supersymmetric (susy) 
sine-Gordon theory. In all these examples, the $N=1$ supersymmetry 
algebra is free of topological charges.
The multicomponent supersymmetric Yang-Lee models are supersymmetric 
versions of the models studied by Freund, Klassen and Melzer (FKM) in 
\cite{FKM}. The latter correspond to the minimal reductions of the 
$A_{2n}^{(2)}$ affine Toda models. We will call their supersymmetric 
extensions the ``susy FKM" models. 

The natural next step after the study of exact bulk
$S$-matrices is to study the same systems in the presence of a boundary.
Again the motivation comes from (open) string theory and the study
of statistical mechanics and condensed matter systems with non-trivial
behavior at a boundary. Examples of the latter are the
various Kondo systems and edge current dynamics in the quantum Hall 
effect.

In this paper we return to the supersymmetric $S$-matrices studied
in \cite{SW,Sch,Ahn1,MS} and study the introduction of a boundary
in those theories. Our strategy will be to assume boundary
conditions (interactions) that preserve both the integrability and
a combination of the left and right supersymmetries.
At the quantum level, these properties imply conditions which will 
allow us to determine the exact reflection matrices. We can
then compare those with predictions based on a purely
classical analysis of supersymmetric boundary conditions,
see, e.g., \cite{IOZ}. 

This paper is organized as follows. 
In section II we discuss the supersymmetric bulk theories 
(including their exact $S$-matrices) that we are going to 
``boundarize", with some special emphasis on the bound state 
structure.
Supersymmetric reflection matrices are introduced in section III.
We study some of their general features and consider the possibility 
of boundary bound states.
In section IV we discuss the bosonic reflection matrices for the
FKM models and for the bound states of the sine-Gordon model. 
Complete supersymmetric reflection matrices are presented
in section V (for the susy FKM models) and VI (for the susy sine-Gordon
model). In VI, we also discuss the relation with the work of
\cite{IOZ}.
In the final section VII we discuss possible extensions and we 
present our conclusions.
 
\section{$S$-matrices with $N=1$ Supersymmetry}

In this section we introduce a ``universal'' bose-fermi 
$S$-matrix $S_{BF}(\theta)$, which is relevant
for theories with $N=1$ supersymmetry without topological
charges. We discuss specific examples, which are
the susy FKM series and the susy sine-Gordon theory.
As a warm-up, we first review some of the basic facts about
particle scattering in integrable supersymmetric theories in
$1+1$ dimensions. Many of the facts and features discussed here 
are shared by general integrable field theories, but to keep 
the discussion short we will introduce those concepts directly 
in the case of supersymmetric models. For more details see 
\cite{SW,Sch,Ahn1}.

\vskip 3mm

\noindent {\underline{\bf{Asymptotic States and the $S$-matrix}}}

The particles in a massive supersymmetric theory are arranged in 
supermultiplets, each having one boson and one fermion $(b_i,f_i)$ 
of equal mass $m_i$, $i=1,\ldots,n$.
The particle operators $A_a(\theta)$ (for bosons or fermions%
\footnote{We shall reserve letters $a,b,a_1,\ldots$ for
particles $b_i$ or $f_i$ and $i,j,i_1,\ldots$ for the
multiplet index $i=1,2,\ldots,n$.})
depend on the rapidity variable $\theta$, which is such that
the on-shell energy and momentum are given by 
$p_0=m_a\cosh(\theta)$ and $p_1=m_a\sinh(\theta)$,
respectively. The in (out) $N$-particle states are written as 
\be
|A_{a_1}(\theta_1)A_{a_2}(\theta_2) \ldots 
A_{a_N}(\theta_N)\rangle_{in(out)} \ , \  \label{b}
\ee
where  $\theta_1 \geq \theta_2 \geq \ldots \geq \theta_N$
for in-states and the other way around for out-states.

Since we are dealing with integrable field theories, we can reduce 
any multiparticle scattering process to two-body scattering
processes. For this reason, the two-body $S$-matrix is a 
key-quantity for understanding the the structure of
the theory. We define it by
\be
S_{a_1a_2}^{b_1b_2}(\theta_1-\theta_2)
|A_{b_2}(\theta_{2})A_{b_1}(\theta_{1})\rangle_{out}
= |A_{a_1}(\theta_{1})A_{a_2}(\theta_{2})\rangle_{in} \ . 
\label{c}
\ee
See figure 1. Notice the inversion of the rapidities for the in and 
out states.

\vskip 0.5cm
\centerline{\epsffile{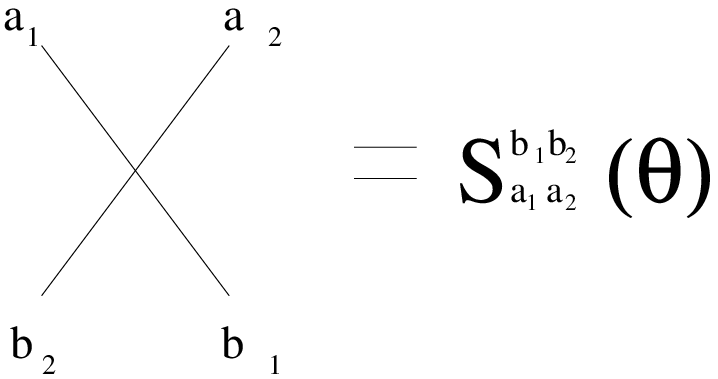}}
\vskip 0.5cm
\centerline{FIG.1 The two-body $S$-matrix element}

In the next subsection we discuss the supersymmetry algebra and the action
of the supercharges on the asymptotic states.
 
\vskip 3mm

\noindent {\underline{\bf{The Supersymmetry Algebra}}}

The supersymmetry of a $d=1+1$ particle theory is expressed by
the presence of conserved supercharges $Q_+(\theta)$ and 
$Q_-(\theta)$ that satisfy
\bea
&& \qquad \qquad \{Q_L,Q_{\pm}\}=0
\nonu
&& Q_+{}^2 = p_0+p_1, \qquad  Q_-{}^2 = p_0-p_1 
\label{abc} \\[2mm]
&& \qquad \qquad \{Q_+,Q_-\}=0 \ .
\nonumber
\eea
The $Q_L$ operator measures the the fermion number of asymptotic 
states. We could have chosen the anti-commutator $\{Q_+,Q_-\}$ 
to be a non-zero c-number $Z$. This would correspond to having a 
topological charge. We shall here restrict our attention to a 
realization of the superalgebra that has $Z=0$. 

The supersymmetry charges $Q_{\pm}(\theta)$ act on asymptotic 
one-particle states according to 
\bea
\begin{array}{ll}
Q_+(\theta)\, |b(\theta) \rangle ={\sqrt{m}} \, e^{\theta \over 2} \,
|f(\theta) \rangle \ ,  \qquad
Q_+(\theta)\, |f(\theta) \rangle ={\sqrt{m}} \, e^{\theta \over 2} \,
|b(\theta) \rangle \\
Q_-(\theta)\, |b(\theta) \rangle = i{\sqrt{m}} \, e^{-{\theta \over 2}} \,
|f(\theta) \rangle \ ,  \qquad
Q_-(\theta)\, |f(\theta) \rangle = -i{\sqrt{m}} \, e^{-{\theta \over 2}}\,
|b(\theta) \rangle \ .
\end{array}
\eea
This corresponds to the following realization
\be
Q_+(\theta)= \sqrt{m} \, e^{\th \over 2}
\left(\begin{array}{cc}
      0 & 1\\
      1 & 0
      \end{array}\right), \quad
Q_-(\theta)= \sqrt{m} \, e^{-{\th \over 2}}
\left(\begin{array}{cc}
      0 & -i\\
      i & 0
      \end{array}\right), \quad 
Q_L(\theta)=
\left(\begin{array}{cc}
      1 & 0\\
      0 & -1
      \end{array}\right)\ \ .
\label{susy4} 
\ee
We have to define the action of the supercharges on multi-particle
states. It is easy to check that the following expression works
\bea
{\cal{Q}}_+(\theta) = \sum_{l=1}^N Q_{+l}(\theta), \qquad \quad
{\cal{Q}}_-(\theta)= \sum_{l=1}^N Q_{-l}(\theta) \ , 
\label{susy2}
\eea
where $Q_{\pm l}(\theta)$ is defined by
\bea
\lefteqn{
Q_{\pm l}(\theta) \, |A_{a_1}(\theta_1) \ldots 
A_{a_N}(\theta_N)\rangle= }
\nonu &&
\prod_{k=1}^{l-1}(-1)^{F_{a_k}} \,
|A_{a_1}(\theta_1)
\ldots A_{a_{l-1}}(\theta_{l-1}) (AQ_{\pm})_{a_l}(\theta_l)
A_{a_{l+1}}(\theta_{l+1})
\ldots A_{a_N}(\theta_N)\rangle \ , \ \label{susy3}
\eea
where $(-1)^{F}$ is $+1$ for a boson and $-1$ for a fermion.

Given this brief description of the supersymmetry algebra and how the
supercharges act on the multi-particle asymptotic states we can move on and
discuss exact $S$-matrices with $N=1$ supersymmetry. We should stress 
again that we have chosen the specific realization (\ref{susy4})
(with zero topological charge) and that this by no means exhausts the 
possible $N=1$ supersymmetric theories.

\vskip 3mm

\noindent{\underline{\bf{The Structure of the $S$-matrix}}}

We concentrate on supersymmetric two-particle $S$-matrices that can be 
written in  the following factorized form
\be
S=S_B \otimes S_{BF} \ , \ \label{d}
\ee
where $S_B$ is the $S$-matrix of the bosonic sector (we shall be 
thinking of a diagonal bosonic $S$-matrix) and $S_{BF}$ is
the supersymmetric piece, responsible for mixing bosons and fermions.
Imposing that this $S$-matrix commutes with the supersymmetry 
charges $Q_{\pm}(\theta)$, one finds that $S_{BF}(\theta)$ gets fixed 
up to one unknown function. Explicitly, one finds the following 
scattering matrix for the particles in the $i$-th and $j$-th 
supermultiplets \cite{Sch}
\be
S_{BF}^{[ij]}(\theta)= f^{[ij]}(\theta)
\left(\begin{array}{cccc}
      1-t \widetilde{t} & 0 & 0 & -i(t+\widetilde{t})\\
      0 & -t+\widetilde{t} & 1+t\widetilde{t} & 0\\
      0 & 1+t\widetilde{t} & t-\widetilde{t} & 0\\
      -i(t+\widetilde{t}) & 0 & 0 & 1-t \widetilde{t}
      \end{array}\right)\ + g^{[ij]}(\theta)
\left(\begin{array}{cccc}
      1 & 0 & 0 & 0 \\
      0 & 1 & 0 & 0 \\
      0 & 0 & 1 & 0 \\
      0 & 0 & 0 & -1
      \end{array}\right)\ \ , \ 
\label{sufa5} 
\ee
where $t=\tanh([\theta+\log(m_{i}/m_{j})]/4)$ and 
$\widetilde{t}=\tanh([\theta-\log(m_{i}/m_{j})]/4)$ and where
we have absorbed a factor of $e^{-{\pi i \over 4}}$ in the definition
of the basic fermionic states $| f_i \rangle$. 
The two functions $f^{[ij]}(\th)$ and $g^{[ij]}(\th)$ are related and 
have to be fixed by the other conditions that the $S$-matrix satisfies. 
The Yang-Baxter
equation fixes the ratio of $f^{[ij]}(\th)$ and $g^{[ij]}(\th)$ up
to a single constant $\alpha$
\be
f^{[ij]}(\theta)={\alpha \over 4i} {\sqrt{m_{i}m_{j}}} 
      \left [
        {2 \cosh(\theta/2)+(\rho^{2} + \rho^{-2})}
        \over {\cosh(\theta/2) \sinh(\theta/2)}
      \right ] g^{[ij]}(\theta)\ , \ \label{sufa6}
\ee
where $\rho=(m_i/m_j)^{1/4}$. 

We have to use analyticity, crossing symmetry and unitarity
to fix the function $g^{[ij]}(\th)$. One expression for
$g^{[ij]}(\th)$ is
\be
 \ g^{[ij]}(\theta)={g_{\Delta_1}(\theta)
                     g_{\Delta_2}(\theta) \over 
                     g_{\Delta_3}(\theta)}\ , \ \label{sufa10}
\ee
where
\be
g_{\Delta}(\theta)={\sinh({\theta \over 2}) \over {\sinh({\theta \over 2}) + 
i \sin({\Delta \pi})}} {\rm  exp}\left(i \int_{0}^{\infty}{dt \over t}
{{\sinh(\Delta t) \sinh((1-\Delta)t)} \over {\cosh^2({t \over 2}) 
\cosh(t)}} \sin({t \theta \over \pi})\right )  \ \label{sufa9}
\ee
and $\Delta_1={1\over2}(i+j)\beta$, $\Delta_2={1\over2}(1 - (i-j)\beta)$ 
and $\Delta_3={1\over2}$. We refer to our previous paper \cite{MS} for 
more details.

We remark that there is an important difference between the periodicity 
properties of diagonal and non-diagonal $S$-matrices. In the diagonal 
case, the conditions for unitarity and crossing symmetry read
\bea
\begin{array}{l}
S_{ij}(\th)S_{ik}(-\th)=\delta_{ik} \ , \qquad
S_{ij}(\th)=S_{i\bar{\jmath}}(i \pi -\th) \ , \ \label{11}
\end{array}
\eea
and these conditions alone imply immediately that $S_{ij}(\th)$ is
$2 \pi i$-periodic. The same is not true for non-diagonal $S$-matrices.
We shall later see that similar remarks apply to the boundary 
reflection matrices.

\vskip 3mm

\noindent{\underline{\bf{Bound State Structure and Three Point Couplings}}

Once we have identified a candidate for an exact $S$-matrix we should 
study the bound state structure of the theory and check the validity of 
the bootstrap principle: the particles that correspond to bound state
poles in the $S$-matrix should be included in the list of asymptotic
particles. This then leads to a number of consistency requirements
that have to be satisfied by the residues at the bound state poles
of the $S$-matrix. For the supersymmetric theories described here one
can derive the following \cite{Sch}.

We denote the 3-point coupling of particles $a$, $b$ and $c$ by 
$f_{ab}^c$. It can be shown that if $f_{ij}^k$ is a non-vanishing
coupling of the bosonic sector (described by $S^{[ij]}_B(\theta)$)
then the full supersymmetric theory has non-zero couplings 
$f_{b_ib_j}^{b_k}$, $f_{b_if_j}^{f_k}$, $f_{f_ib_j}^{f_k}$ and
$f_{f_if_j}^{b_k}$, satisfying
\be
{{f_{f_if_j}^{b_k}} \over {f_{b_ib_j}^{b_k}}}=\left({{m_i+m_j-m_k} \over
{m_i+m_j+m_k}} \right)^{1 \over 2} \ . \ \label{bs}
\label{ratioff}
\ee
Furthermore, the value of the free parameter $\alpha$ in the bose-fermi 
$S$-matrix (see (\ref{sufa6})) can be expressed in terms of the masses
$m_i$, $m_j$ and $m_k$ of any three multiplets $i,j,k$ with non-zero
3-point coupling $f_{ij}^k$, according to 
\be
\alpha= -{{(2m_i^2m_j^2+2m_i^2m_k^2+2m_j^2m_k^2-m_i^4-m_j^4-m_k^4)^{1\over2}} 
    \over {2m_im_jm_k}} \ . \ \label{bs1}
\ee
Note that one free constant $\alpha$ has to fit all non-zero couplings 
of the bosonic theory. Clearly, this will only be possible in very special 
cases. This then shows the type of supersymmetrization that is described 
here is only possible for a selected set of bosonic theories.

If the multiplets $i$, $j$ and $k$ have a non-vanishing
coupling, the two-body $S$-matrix 
$S^{[ij]}(\theta)$ will have a pole at
$\th=iu_{ij}^k$ where $u_{ij}^k$ satisfies
\be
\cos(u_{ij}^k)={{m_k^2-m_i^2-m_j^2} \over {2m_im_j}} \ . \ \label{bs2}
\ee
Clearly, the $u_{ij}^k,u_{jk}^i,u_{ki}^j$ are the external
angles of a triangle with sides $m_i$, $m_j$ and $m_k$ (and so we define 
the internal angles $\overline{u}_{ij}^k=\pi-u_{ij}^k$).

\vskip 3mm

\noindent {\underline{\bf{Examples: Supersymmetric FKM Series
                          and Supersymmetric sine-Gordon}}}

It was observed in \cite{Sch} that the conditions (\ref{bs1})
are all satisfied in theories with a mass spectrum 
\be
m_j={\sin(j\beta \pi) \over {\sin(\beta \pi)}} \ , \qquad
  j=1,2,\ldots \ \label{a}
\ee
and the following allowed fusion pattern: $f_{ij}^k$ is non-vanishing 
if $k=|i-j|$ or $k=i+j$. In this case $\alpha=-\sin(\beta \pi)$.

This combination of spectrum and fusion rules is known to be 
realized in at least two (series of) examples: the sine-Gordon 
theory and the multicomponent Yang-Lee (or FKM \cite{FKM}) minimal 
models.
In the sine-Gordon case, the above are the masses and 
fusion rules of the breathers (bound-state) and $\beta$ is 
related to the (adjustable) coupling constant $\beta_{sG}$. The FKM 
minimal models correspond to $\beta_{sG}=1/(2n+1)$, $n=1,2,\ldots$. They 
can to some extent be viewed as consistent truncations of the 
sine-Gordon bound-state sectors. In the FKM models there are additional
non-zero couplings $f_{ij}^k$ for $i+j+k=2n+1$ (again, these are
consistent with the conditions (\ref{bs1})).

One thus expects that the scattering matrices for the $N=1$ 
supersymmetric extensions of both the FKM (susy FKM) and the 
sine-Gordon (susy sine-Gordon) theories will be of the form described 
above. This has been confirmed in \cite{Sch,Ahn2,MS,SW,Ahn1}. In the 
remainder of this section we briefly summarize these results.

The susy FKM models are integrable deformations of specific 
non-unitary superconformal minimal models, labeled 
by an integer $n=1,2,\ldots$, and of central charge 
given by $c_n=-3n(4n+3)/(2n+2)$.
They are formally defined through a perturbed superconformal
field theory with action
\be
S_{\lambda}=
S+\lambda \int\overline{G}_{-{1 \over 2}} G_{-{1\over 2}} \phi_{h,h}
d^2x \ , \ \label{a1}
\ee
where $\phi_{h,h}$ is a primary field in the Neveu-Schwarz sector
of the left and right chiral superconformal algebras and $h=h_{(1,3)}$.
This is an integrable perturbation that is manifestly supersymmetric.
Having introduced a relevant parameter $\lambda$, we obtain a
massive theory, with particles $(b_j,f_j)$, $j=1,2,\ldots,n$, with
masses as in (\ref{a}) with $\beta=1/(2n+1)$. 
The $S$-matrix for these theories takes the form (\ref{d})
with $S_B$ given by the $S$-matrix of the bosonic multi-component
Yang-Lee model as found by Freund, Klassen and Melzer (FKM) in 
\cite{FKM} 
\be
S^{[ij]}_B(\theta)=F_{|i-j|\beta}(\theta) \left[ 
  F_{(|i-j|+2)\beta}(\theta) \ldots
  F_{(i+j-2)\beta}(\theta) \right]^2 F_{(i+j)\beta}(\theta)\ , \ 
\label{sufa4}
\ee
with $F_{\alpha}(\theta)={{\sinh(\theta)+i\sin(\alpha \pi)} \over 
{\sinh(\theta)-i\sin(\alpha \pi)}}$ and $\alpha=-\sin(\pi/(2n+1))$.
These bosonic $S$-matrices
can be viewed as `minimal reductions' of the $S$-matrices 
of the twisted affine Toda theories based on $A_{2n}^{(2)}$. 

The first model ($n=1$) in the susy FKM series is a supersymmetric
version of the perturbed Yang-Lee conformal field theory. It has a 
single massive supermultiplet $(b,f)$ with non-zero self-couplings
$f_{bb}^b = \sqrt{3} \, f_{ff}^b$ that correspond to a 
pole at $\theta= i{2\pi \over 3}$ in the bosonic factor $S_B(\theta)$
of the $S$-matrix.
This theory may justifiably be called the ``world's simplest 
interesting supersymmetric scattering theory'', and as such it 
serves as a prototype for theoretical investigations of more 
complicated supersymmetric models.

The supersymmetric sine-Gordon model is defined
by the following action in Euclidean space-time
\bea
S_{ssG}=\int_{-\infty}^{\infty}dy\int_{-\infty}^{\infty}dx  
&&\left\{ \STR {1\over2}(\partial_x\phi)^2+{1\over2}(\partial_y\phi)^2-
{\bar{\psi}}(\partial_x-i\partial_y){\bar{\psi}}+
\psi(\partial_x+i\partial_y)\psi-\right.\nonu
&&\left.-{{m^2}\over{\beta_{ssG}^2}}\cos(\beta_{ssG}\phi)-
2m{\bar{\psi}}\psi\cos({{\beta_{ssG}\phi}\over2}) \STR \right\} , 
\label{ssG1}
\eea
where $\phi$ is the bosonic field and $\psi$ and $\bar{\psi}$ are the
components of a Majorana fermion. The spectrum of the full quantum theory 
contains (anti-)soliton multiplets and bound state multiplets
$(b_j,f_j)$, $j=1,2,\ldots < \lambda$, $\lambda = 
2\pi \left( 1 - (\beta_{ssG}^2 / 4\pi) \right)/\beta_{ssG}^2 $,
of masses (\ref{a}) with $\beta= 1/(2\lambda)$.
The $S$-matrix for the bound state multiplets \cite{SW,Ahn1}
again takes the factorized form (\ref{d}), where the $S_B(\theta)$ 
are now the breather $S$-matrices of the bosonic sine-Gordon model. 
They can be found in the original paper \cite{ZZ}.

The truncation from susy sine-Gordon to a minimal susy FKM
model is possible when $\lambda$ is of the form 
$\lambda={2n+1 \over 2}$. Consider for example $n=1$, which gives 
$\lambda = {3 \over 2}$. At this coupling, only the lightest 
breather multiplet ($j=1$) is stable. However, it so happens that 
the second breather multiplet, which occurs as a virtual bound state 
in the $S$-matrix of the first multiplet ($f_{11}^2\neq 0$), 
has the same mass as the first one. It is then consistent to 
consider the minimal reduction, where the $j=1,2$ bound state 
multiplets are `folded' onto a single particle with self-coupling
($f_{11}^1 \neq 0$), and this folded theory is essentially
the same as the Yang-Lee scattering theory. This formal connection 
explains the similarity between the $S$-matrices for the susy 
sine-Gordon and susy FKM theories.

\section{Supersymmetric Reflection Matrices: Generalities}

Once we have understood the bulk $S$-matrix for a physical
theory it is natural to study the same theory in the presence 
of a boundary.
The reasons for studying boundary theories are both theoretical 
and physical. From the theoretical point of view it is an 
interesting (and difficult) problem to find and identify the boundary 
reflection matrices that correspond to given boundary conditions 
or a given boundary action. From the practical point of view,
many physical systems such as the Kondo system and edge excitations 
in the fractional quantum Hall effect correspond to boundary 
problems.

We should remark that conserved quantities in the bulk may be no
longer be conserved after the introduction of a boundary, as is for 
example the case for linear momentum. So we expect that not all 
boundary actions lead to integrable boundary theories and if
we insist on integrability we shall have to be careful about which
boundary action we are picking. In this paper we shall focus
on boundary conditions that preserve both the integrability
and the supersymmetry of the bulk theory.

In order to have the complete description of a particle theory in 
the presence of a boundary we have to understand how particles scatter 
off that boundary. In an integrable theory one expects that this
scattering is one-to-one. The corresponding amplitudes  
are contained in a reflection matrix $R^b_a(\theta)$. This amplitude is
shown in figure 2. In this section 
we review some general features of such reflection matrices and discuss 
the implementation of supersymmetry. Our presentation follows the 
spirit and the notations of \cite{GZ}. We refer to \cite{Wa} for a 
study of $N=2$ supersymmetry in integrable models with a boundary.

\vskip 0.5cm
\centerline{\epsffile{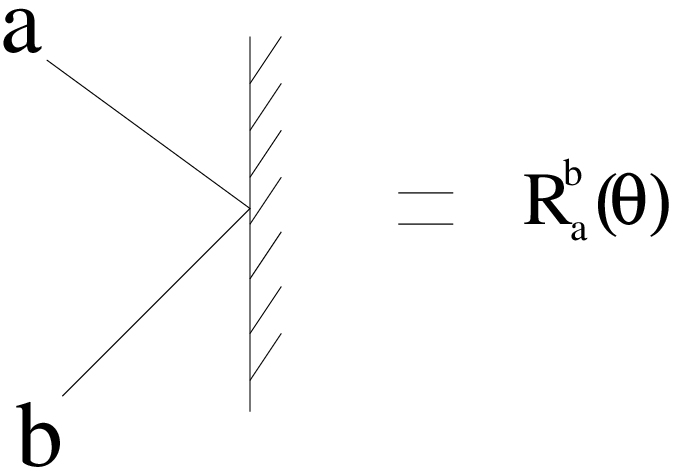}}
\vskip 0.5cm
\centerline{FIG.2 The reflection matrix}

We will restrict our discussion to the case where the boundary has 
no structure, so that the reflection matrix can be assumed
to be diagonal, $R_a^b(\th)=\delta_a^b R_a(\th)$, no sum over $a$. 
(In the presence of boundary bound states more general reflection
matrices should be considered, see below.) The one-particle
reflection ampitudes are subject to a number of conditions
that are most easily understood by drawing some pictures,
see \cite{FK1,GZ}. In an analogous way as for the bulk theories, 
we have boundary Yang-Baxter equations (BYBE) (see figure 3)
\footnote{no sum over $a_1,a_2,b_1,b_2$},
\bea
\lefteqn{
R_{a_2}(\th_2)
S_{a_1a_2}^{c_1d_2}(\th_1+\th_2)
R_{c_1}(\th_1)
S_{d_2c_1}^{b_2b_1}(\th_1-\th_2)=}
\nonu
&& \qquad 
S_{a_1a_2}^{c_1c_2}(\th_1-\th_2)
R_{c_1}(\th_1)
S_{c_2c_1}^{b_2b_1}(\th_1+\th_2)
R_{b_2}(\th_2) \ . 
\label{bdr}
\eea

\vskip 0.5cm
\centerline{\epsffile{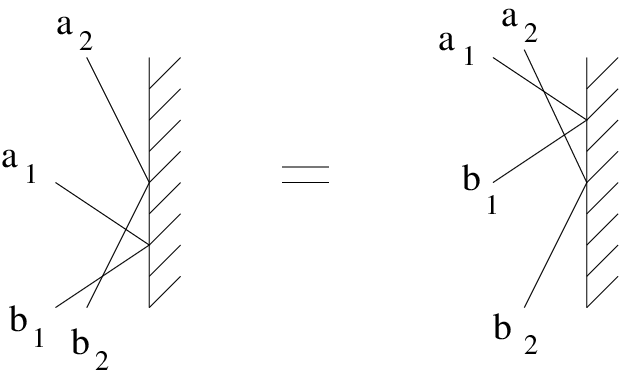}}
\vskip 0.5cm
\centerline{FIG.3 The boundary Yang-Baxter equation}

We have also a unitarity condition for the reflection matrix,
\be
R_a(\theta)R_a(-\theta)=1 \ , \ \label{bdr3}
\ee
and a condition called boundary crossing-unitarity 
\be
R_a(i {\pi \over 2}-\theta)= S^{aa}_{bb}(\theta)
R_b(i {\pi \over 2}+\theta) \ , \ \label{bdr2}
\ee
where we sum over $b$. If the bulk $S$-matrix has non-trivial bound 
state poles there are additional ``boundary bootstrap'' conditions, which
include the identity \footnote{summing over $a_1$ and $b_1$}
\be
f^{ab}_c R_c(\theta)=
f^{a_1b_1}_c
R_b(\theta-i{\overline{u}}_{bc}^a) 
S_{b_1a_1}^{ba}(2\theta+i{\overline{u}}_{ac}^b-i{\overline{u}}_{bc}^a)
R_{a_1}(\theta+i{\overline{u}}_{ac}^b)
\ , \ \label{bdr4}
\ee
where  $f^{ab}_c$ are the fusion constants discussed before
and we used the fact that we are dealing with neutral particles, 
$\bar{a}=a$. See figure 4.

\vskip 0.5cm
\centerline{\epsffile{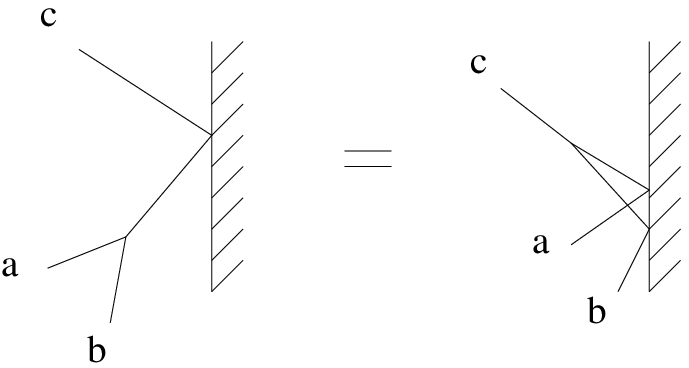}}
\vskip 0.5cm
\centerline{FIG.4 The boundary bootstrap condition}

The usual strategy for handling these equations, given the 
knowledge of a bulk $S$-matrix, is be to first solve the BYBE and 
then take care of the other equations and of the boundary bootstrap 
requirements. However, in our case we have supersymmetry as an 
additional ingredient which we can use to simplify the analysis.

We shall assume that the reflection matrix has a factorized structure 
similar to that of the bulk $S$-matrix, namely
\be
R(\theta)=R_B(\theta) \otimes R_{BF}(\theta) \ , \ \label{Ia}
\ee
where $R_B(\theta)$ is the reflection matrix for the bosonic part of the
theory, and $R_{BF}(\theta)$ is the ``supersymmetric" part of the 
reflection matrix. Given this form for the reflection matrix, all bosonic
equations will factor out and we have therefore two problems to
consider. The first is to find the reflection matrices $R_B(\theta)$
corresponding to the bosonic sector. In the cases at hand these will be 
reflection matrices for the FKM series (which are for the first time 
given in this paper) and the reflection matrices for the sine-Gordon 
breathers, which were given by Ghoshal in \cite{Gh}. The second step will
be to add a factor $R_{BF}(\theta)$ that describes the relative
amplitudes for bosons and fermions when scattering off the boundary.
The factor $R_{BF}(\theta)$ is subject to a non-trivial BYBE. Our
strategy will be to derive $R_{BF}(\theta)$ by imposing supersymmetry
and to check that the resulting expression indeed satisfies the 
BYBE.

In the presence of bound states (bulk or boundary), there are 
consistency requirements that are not automatically satisfied 
by the factorized expression (\ref{Ia}). These will be discussed 
below.

\vskip 3mm

\noindent{\underline{\bf Boundary Supersymmetry}}

In the presence of a boundary, only a specific linear combination
of the left and right supersymmetries can be preserved. This is
of course familiar from open string theory, see also \cite{Wa}.
The invariant combination should satisfy the ``commutation''
relation
\be
Q(\theta) R(\theta) = R(\theta)Q(-\theta) \ , \ \label{Ib}
\ee
where $Q(\theta) = a \, Q_+(\theta) + b \, Q_-(\theta)$,
and $a$ and $b$ are constants. It is easy to see that the only solutions 
to these equation are simply
\be
Q^{(\pm)}(\theta)= Q_+(\theta) \mp Q_-(\theta) \ , \ \label{Ic}
\ee
together with 
\be
R^{(\pm)}_{BF}(\theta)=Z^{(\pm)}(\theta)
\left(\begin{array}{cc}
     \cosh({\theta \over 2}\pm i{\pi \over 4}) & e^{i{\pi \over 4}}Y(\th)\\
       e^{-i{\pi \over 4}}Y(\th) & \cosh({\theta \over 2}\mp i{\pi \over 4})
      \end{array} \right)\ \ . \ \label{e2}
\ee
These are thus the most general reflection matrices compatible with our 
realization of $N=1$ supersymmetry (which assumed the absence of
topological charges). 
If we impose now that we have a boundary with no structure we should set 
$Y(\th)=0$, since $Y(\th)$ is (up to a phase) the amplitude of a scattering 
off the boundary that changes the fermion number. The result is then
\be
R^{(\pm)}_{BF}(\theta)=Z^{(\pm)}(\theta)
\left(\begin{array}{cc}
     \cosh({\theta \over 2}\pm i{\pi \over 4}) & 0\\
       0 & \cosh({\theta \over 2}\mp i{\pi \over 4})
      \end{array} \right)\ \ . \ \label{e3}
\ee
It is surprising that we have been able to determine the form of the
possible supersymmetry charges and the structure of the reflection matrix 
in one go. The result is that the amplitudes for a particle and its 
superpartner scattering off the boundary are related in a universal way 
(independent of masses) by
\be
{R_b^{(\pm)}(\theta) \over R_f^{(\pm)}(\theta)} =
{{\cosh({\theta \over 2} \pm i {\pi \over 4})} \over 
 {\cosh({\theta \over 2} \mp i {\pi \over 4})}} \ . \ \label{Id}
\ee

It is important to stress that this result assumes that we started with
an integrable supersymmetric model and introduced a boundary that
preserves {\it both} integrability and supersymmetry. We just learned 
that these conditions are very restrictive and in fact leave no free
parameters (except for a choice of sign) in the choice of boundary 
conditions. A similar conclusion has been reached in a purely
classical analysis of the susy sine-Gordon theory
\cite{IOZ}. In section VI we shall propose a precise connection 
between our results and the work of \cite{IOZ}.

In order to fix the prefactor $Z^{(\pm)}(\theta)$ we have to use 
the unitarity and boundary crossing-unitarity conditions. They lead to 
\bea
Z_j^{(\pm)}(\theta)Z_j^{(\pm)}(-\theta)&=&
{2 \over {\cosh(\theta)}}\ , 
\nonu
{Z_j^{(\pm)}(\ipi-\theta) \over {Z_j^{(\pm)}(\ipi+\theta)}}&=&
\mp S_{b_jb_j}^{b_jb_j} (2\theta)+ if^{(\pm)}(\theta)
S_{f_jf_j}^{b_jb_j}(2\theta) \ , \ \label{bdr11} 
\eea
where $f^{(+)}(\theta)=\coth({\theta \over 2})$ and $f^{(-)}(\theta)=
\tanh({\theta \over 2})$, depending on which sign we pick in
$R_{BF}^{(\pm)}$. The index $j$ in the notation $Z^{(\pm)}_j(\theta)$ 
expresses the dependence of these functions on the mass $m_j$ of the
reflecting particle. 

In section V we shall present explicit solutions to
the equations (\ref{bdr11}). Such solutions are 
unique up to multiplicative factors $\phi(\theta)$ 
satisfying
\be
\phi(\theta)\phi(-\theta) = 1 \ , \qquad
{\phi(i {\pi \over2 } + \theta) \over
\phi(i {\pi \over2 } - \theta)} = 1 \ .
\ee
Such factors are the boundary analogues of the familiar
CDD factors.

\noindent{\underline{\bf{The Free Case}}}

We can apply the preceding discussion to an extremely simple, 
but nonetheless useful, example: the free supersymmetric theory 
with one boson and one fermion. In this case we have $S_B(\th)=1$ 
and the non-vanishing entries of $S_{BF}(\theta)$ are given by 
$S_{bb}^{bb}=1$, $S_{ff}^{ff}=-1$, and $S_{bf}^{bf} = S_{fb}^{fb}=1$. 
Clearly, there are no bound states that we should worry about. 
Assuming the trivial amplitude $R_b(\th)=1$, we find
\be
R^{(\pm)}(\theta)=
\left(\begin{array}{cc}
      1 & 0\\
      0 &{{\cosh({\theta \over 2}\mp i{\pi \over 4})} \over
          {\cosh({\theta \over 2}\pm i{\pi \over 4})}}
      \end{array} \right)\ \ . 
\label{ee}
\ee
Comparing with the result by Ghoshal and Zamolodchikov \cite{GZ} for
the free Majorana fermion, we see that $R^{(+)}(\th)$ corresponds to the
free boundary condition ($\psi-{\bar{\psi}}=0$) and $R^{(-)}(\th)$
to the fixed boundary condition ($\psi+{\bar{\psi}}=0$). 

\vskip 3mm

\noindent{\underline{\bf{Boundary Bootstrap and Boundary Bound States}}}

The presence of poles (with imaginary part between $0$ and $\pi/2$) 
in a boundary reflection matrix signals a resonance, where bulk
particles can `merge' with the boundary state. 

A resonance at rapidity 
$i {\pi \over 2}$, i.e., at zero energy, indicates that the (unique)
boundary state $|0\rangle_B$ has a non-zero coupling $g^c$ with
zero-energy bulk particles of type $c$. If there is a non-zero bulk
coupling $f_{ab}^c$ with $m_a=m_b$, one expects a pole of $R^b_a(\th)$ 
at $\th=i{{{\overline{u}}_{ab}^c} \over 2}$ with
\be
R^b_a(\th) \sim - {i \over 2} {{f_{ab}^{c}\, g^c} \over 
{\th -i{{\overline{u}_{ab}^c} \over 2}}} \ , \ \label{bss}
\ee
as $\th \rightarrow i{{{\overline{u}}_{ab}^c} \over 2}$.

In the presence of $N=1$ supersymmetry, one would in 
principle expect that the bound state couplings $g^i$
branch into non-zero $g^{b_i}$ and $g^{f_i}$. However, in view
of (\ref{bss}), a non-zero value for $g^{f_i}$ would
imply singularities in amplitudes that describe reflection
processes that violate fermion number. Since we assumed
that such processes do not occur, we conclude that
all fermionic couplings $g^{f_i}$ should vanish. This then
implies that the ratio $R_f/R_b$ should have a 
zero at $\theta = i{\pi \over 2}$, which is the
case for $R_{BF}^{(-)}$ but not for $R_{BF}^{(+)}$.
We conclude that in the situation with non-vanishing
boundary couplings $g^i$, the supersymmetrization
can only be done with the factor $R_{BF}^{(-)}$.

If we now assume a non-vanishing coupling among
the particles $(b_i,b_i,b_j)$, we derive that 
\bea
R_{b_i}(\th)\sim - {i \over 2} {{f_{b_ib_i}^{b_j}g^{b_j}} \over 
{\th -i{{\overline{u}_{ii}^j} \over 2}}}\ , \qquad 
R_{f_i}(\th)\sim - {i \over 2} {{f_{f_if_i}^{b_j}g^{b_j}} \over 
{\th -i{{\overline{u}_{ii}^j} \over 2}}} \ , \ \label{bss1}
\eea
and taking the ratio we obtain
\be
{{R_{b_i}(i{{\overline{u}_{ii}^j} \over 2})} \over
 {R_{f_i}(i{{\overline{u}_{ii}^j} \over 2})}}=
{{f_{b_ib_i}^{b_j}} \over {f_{f_if_i}^{b_j}}} \ . \ \label{bss3}
\ee
The RHS of this equation is a feature of the bulk scattering
theory, and it depends solely on the masses $m_i$ and $m_j$
(see (\ref{ratioff})). The LHS (which does not depend on the 
normalization $Z^{(\pm)}(\th)$), is precisely the universal 
ratio (\ref{Id}) (based on $R_{BF}^{(-)}$) of the boson and 
fermion reflection amplitudes. Thus, while the value of the RHS 
is a consequence of the bulk supersymmetry, the LHS has been dictated 
by boundary supersymmetry. The fact that these ratios do indeed 
agree (as is easily worked out) is therefore a nice confirmation 
of the consistency of our description.

If the boundary reflection amplitudes have resonances that
can not be taken into account by the above, this may signal
the presence of `boundary bound states'. These are stable 
configurations in which the boundary can exist and which can 
be excited by incident bulk particles. A boundary 
bound state $|\alpha\rangle_B$ of energy $e_{\alpha}$ can be 
excited by an incident particle of mass $m$ and rapidity 
$\theta = i v_{0a}^{\alpha}$, provided
\be
e_0+m_a\cos(v_{0a}^{\alpha})=e_{\alpha} \ . \ \label{e5}
\ee
The behavior of $R_a^b(\th)$ is then given by
\be
R_a^b(\th) \sim {i \over 2} { g_{a0}^{\alpha} g^{b0}_{\alpha} 
           \over \th -iv_{0a}^{\alpha} }
\ , \ \label{e6}
\ee
as $\th \rightarrow iv_{0a}^{\alpha}$. In the presence of 
supersymmetry, a boundary coupling $g_{a0}^{\alpha}$
will branch into a coupling $g_{b0}^{\beta}$ for the bosons
and a coupling $g_{f0}^{\varphi}$ for the fermions. In
other words, there are boundary states $|\beta\rangle_B$
and $|\varphi\rangle_B$, which form a multiplet under
$N=1$ boundary supersymmetry. The relative strength of
$g_{b0}^{\beta}$ and $g_{f0}^{\varphi}$ can for example
be derived from the observation that the residue
of $R_b(\theta)$ ($R_f(\theta)$) at the pole
$\theta \rightarrow i v_{bo}^\beta$ is proportional to
$(g_{b0}^\beta)^2$ ( $(g_{f0}^\varphi)^2)$. This implies
that
\be
\left({ {g_{b0}^{\beta}} \over {g_{f0}^{\varphi}} }
 \right)_{(\pm)}^2 
= { \cosh({iv \over 2} \pm i {\pi \over 4}) \over
    \cosh({iv \over 2} \mp i {\pi \over 4}) }
\label{ratiogg}
\ee
with $v=v_{b0}^\beta$. 

We can formally write the excited boundary states as
\be
\lim_{\epsilon \rightarrow 0} \, \epsilon \, b(iv+\epsilon) \, |0\rangle_B
= g_{b0}^\beta \, |\beta\rangle_B \ , \qquad
\lim_{\epsilon \rightarrow 0} \, \epsilon \, f(iv+\epsilon) \, |0\rangle_B
= g_{f0}^\varphi \, |\varphi\rangle_B \ .
\ee
Using (\ref{ratiogg}) we can work out the action of
boundary supersymmetry on these states, with the result
\bea
& Q^{(+)} |\beta \rangle_B = \sqrt{2m \cos v} \, |\varphi \rangle_B \ ,
& \qquad
Q^{(+)} |\varphi \rangle_B = \sqrt{2m \cos v} \, |\beta \rangle_B \ ,
\nonu
& Q^{(-)} |\beta \rangle_B = i \sqrt{2m \cos v} \, |\varphi \rangle_B \ ,
& \qquad
Q^{(-)} |\varphi \rangle_B = -i \sqrt{2m \cos v} \, |\beta \rangle_B \ .
\eea
We see that $[Q^{(\pm)}]^2 = 2m \cos v = 2 (e_\beta - e_0)$ as it 
should be.

Focusing on the case with $R_{BF}^{(-)}$, we can imagine
starting from the situation with $v=\pi/2$ and then moving
$v$ down along the imaginary axis. In the supersymmetric
theory, this process represents a kind of supersymmetry
breaking at the boundary: at $v=\pi/2$ the boundary state 
has zero energy and it is annihilated by supersymmetry,
while for $v < \pi/2$ the energy is positive and 
supersymmetry no longer annihilates the state.

In the presence of boundary bound states, one may consider 
more general reflection amplitudes, such as $R_{a\alpha}^{b\beta}$, 
which describes a process where a particle $a(\theta)$ reflects into 
$b(-\theta)$ while the boundary makes a transition from 
$|\alpha\rangle_B$ to $|\beta\rangle_B$. Clearly, our formalism can 
be extended to amplitudes of this type, but we shall not do so in
this paper.

As we will see later, not all bosonic reflection matrices can be 
supersymmetrized, even though we have a general form for $R_{BF}(\th)$.
This can probably be understood through the analysis of residue 
conditions for more general reflection amplitudes, such as 
$R_{a\alpha}^{b\beta}$, which involve boundary bound states \cite{Wa}. 
In the bulk theories, such an analysis \cite{Sch} has led to the 
condition (\ref{bs1}), which shows that many bosonic theories simply 
can not be supersymmetrized by adding a factor $S_{BF}$ to their 
bosonic $S$-matrix $S_B$. In a similar way, we expect that the product 
form $R=R_B \otimes R_{BF}$ for a supersymmetric reflection matrix will 
only be possible in theories with a specific and restricted set of 
boundary bound states.

\section{Bosonic Reflection Matrices}

As we have discussed earlier, we need to find the reflection matrices
for the bosonic sectors of the models we are studying. The reflection
matrices for the FKM series (which are minimal reductions of the
$A_{2n}^{(2)}$ affine Toda theories) have been studied for a few
special cases only \cite{GZ,FK1}. Here we shall give two solutions
for general $n=1,2,\ldots$. The reflection matrices for bound states 
of the boundary sine-Gordon theory were studied by Ghoshal in \cite{Gh};
for completeness we quote his results below. We shall also comment on 
the relation between the reflection matrices for sine-Gordon
breathers and those for the FKM models.

\vskip 3mm

\noindent{\underline{\bf{The FKM Series}}}

In this subsection we derive two possible reflection matrices
for each of the bosonic FKM models.

We begin our derivation by briefly reviewing the work of
Fring and K\"oberle \cite{FK1,FK2} on the reflection matrices for 
the affine Toda field theories based on the untwisted
affine algebras $A_l^{(1)}$. These are massive, integrable
QFT's, with particles of masses $m_j$, $j=1,2,\ldots,l$,
given by the formula (\ref{a}) with $\beta=l+1$.
Starting from a ``block" structure for the $S$-matrix, 
Fring and K\"oberle were able to conjecture the following form of 
the reflection matrices 
\be
R^j_{A_l^{(1)}}(\theta)= \prod_{i=1}^{\mu(j)}
{\cal{W}}_{l-2\mu(j)+2\nu(i)}(\theta) \ , \ \label{kfr}
\ee
with $\mu(i)$ given by
\bea
\mu(i)=\left\{
\begin{array}{ll}
i   & \qquad \hbox{for} \;  i \leq [\frac{h}{2}] \\
h-i & \qquad \hbox{for} \;  i > [\frac{h}{2}]  \;\; ,
\end{array}    \right.
\eea
and $\nu(i)$ by
\bea
\nu(i) = \left\{
\begin{array}{ll}
i   & \qquad \hbox{for} \;  i \; \hbox{odd} \\
i+h & \qquad \hbox{for} \;  i \; \hbox{even}  \;\; ,
\end{array}    \right.
\eea
where $h$ is the Coxeter number associated to the affine Lie algebra
$A_l^{(1)}$, $h=l+1$. The precise expression for the building block
${\cal{W}}_x(\theta)$, which may be found in \cite{FK2}, of course 
depends on the value of the coupling constant of the affine Toda theory.
The expressions (\ref{kfr}) are free of poles in the physical strip.

We shall take this result as a starting point for deriving the
reflection matrices for FKM series of minimal models. To get
there, we need to make two steps. The first step is to reduce
the result for the $A_l^{(1)}$ Toda series to parameter-free 
`minimal' reflection matrices. These then describe the 
`minimal' boundary scattering of a set of $l$ particles, whose bulk 
$S$-matrix was first given in \cite{KSw}. This minimal
scattering theory describes the IR behavior of an integrable 
perturbation of a CFT of central charge $c_l = {2l \over l+3}$.
In the above formulas, the reduction to the minimal case
can be done by replacing the ${\cal W}_x(\th)$ used in \cite{FK2}
by the following parameter-free expression
\be
W_x(\theta)=
{\sinh{1 \over 2}(\theta+i{\pi(1-x-h) \over {2h}})\over 
 \sinh{1 \over 2}(\theta-i{\pi(1-x-h) \over {2h}})} 
{\sinh{1 \over 2}(\theta-i{\pi(1+x+h) \over {2h}})\over
 \sinh{1 \over 2}(\theta+i{\pi(1+x+h) \over {2h}})} \ . 
\label{KFW}
\ee

The second step is to apply a procedure called `folding in
half' to the minimal reduction of the $A_l^{(1)}$ Toda result 
with $l=2n$. It is well-known that the bulk $S$-matrices
for the minimal reductions of the $A_{2n}^{(1)}$ and $A_{2n}^{(2)}$ 
affine Toda theories satisfy the relation
\be
S_{ij}^{A_{2n}^{(2)}}(\theta) = 
S_{ij}^{A_{2n}^{(1)}}(\theta) \,  
S_{i\bar\jmath}^{A_{2n}^{(1)}}(\theta) \ 
\ee 
with $i,j=1,2,\ldots,n$ and $\bar\jmath = (h-j)$. 
This folding relation
has interesting consequences at the level
of the TBA based on these $S$-matrices. In particular, it
follows that the effective central charge of the minimal
$A_{2n}^{(2)}$ theory is precisely half that of the
theory based on $A_{2n}^{(1)}$. The folding relation of
the bulk $S$-matrices suggests a similar structure for the
reflection matrices, where we expect that every factor
$W_x(\th)$ will be multiplied
by a factor $W_{h-x}(\th)$. This then is up to possible shifts
of the indices $x$ by the amount $2h=4n+2$. These
we have determined `by hand', by insisting on a consistent
pole structure and on the boundary bootstrap conditions.

We claim that the following two series of reflection matrices
are consistent solutions for the FKM minimal theories
\bea
&& R_{(1)}^j (\th)= \prod_{i=1}^j \left[ W_{2n-2j + 2\nu(i)}(\th) \,
                                 W_{-4n+2j-1 - 2i }(\th) \right] \ ,
\nonu
&& R_{(2)}^j (\th)= \prod_{i=1}^j \left[ W_{2n-2j + 2\nu(i)}(\th) \,
                                 W_{2j+1 -2i}(\th) \right] \ .
\label{r1r2}
\eea
We can check this conjecture against a few known cases. Ghoshal and 
Zamolodchikov studied the Yang-Lee model (the $n=1$ model in
the FKM series) in \cite{GZ}. For $n=1$ (\ref{r1r2}) gives
\be
R_{(1)}^1 (\th)= W_2 (\th) W_{-5}(\th)\ , \quad R_{(2)}^1(\th) = 
W_2(\th) W_1(\th) \ ,
\ee
in agreement with the result of \cite{GZ}%
\footnote{There is a sign mistake in the equation (3.51) of \cite{GZ}.}.
For $n=2$ the result is%
\footnote{${\cal{W}}_4(\th)$  should be replaced by
     ${\cal{W}}_{-6}(\th)$ for the $A_4^{(2)}$ in Fring and 
     K\"oberle's \cite{FK1} equation (5.81).}
\bea &&
R_{(1)}^1(\th) = W_4(\th) W_{-9}(\th)\ , \quad
R_{(1)}^2(\th) = \left[ W_2(\th) W_{-7}(\th) \right]
                 \left[ W_{-6}(\th) W_{-9}(\th) \right] \ ,  
\nonu &&
R_{(2)}^1(\th) = W_4(\th) W_1(\th) \ , \quad 
R_{(2)}^2(\th) = \left[ W_2(\th) W_3(\th) \right]
                 \left[ W_{-6}(\th) W_{1}(\th) \right] \ .
\label{eq} 
\eea

In general, the solution $R_{(2)}^j(\th)$ is obtained
from $R_{(1)}^j(\th)$ by the replacement $W_x(\th) \rightarrow 
W_{x+2h}(\th)$ for all odd $x$. 
It can easily be checked that this replacement
corresponds to a multiplication by a CDD factor that is consistent
with the boundary bootstrap equations.

The reflection amplitude $R_{(1)}^1(\theta)$ for the lightest 
particle has `physical' poles at $\theta = i{(2n-1)\pi \over 2(2n+1)}$,
and $\theta = i{\pi \over2}$.
The first of these is at $\theta=i {\overline{u}_{11}^2 \over 2}$ and 
corresponds to a nonzero boundary coupling $g^2$. As we have seen, the pole
at $\theta = i {\pi \over 2}$ indicates a non-zero value
of $g^1$. The reflection amplitude $R_{(2)}^1(\theta)$ is free
of poles in the physical strip.

\vskip 3mm

\newpage

\noindent{\underline{\bf{The sine-Gordon Bound States}}}

The boundary sine-Gordon model was studied in \cite{GZ,Gh}, 
where these authors found the reflection matrices for solitons 
and breathers. We shall need the latter ones. Most importantly, 
it was found that the sine-Gordon reflection matrices have
two free parameters, called $\eta$ and $\vartheta$. These
parameters are believed to correspond to free parameters
$M$ and $\phi_0$ in the most general boundary action that
preserves the integrability of the sine-Gordon theory.

Following Ghoshal \cite{Gh}, we write the reflection matrix
for the $j$-th breather as  
$R_{sG}^j(\th|\eta,\vartheta) = R_0^j(\th) \, R_1^j(\th)$, where
\bea
R_0^j(\th)=(-1)^{j+1}
{{\cosh(\qh +i{{j \pi} \over {4 \lambda}})
  \cosh(\qh -i{\pi \over 4}-i{{j \pi} \over {4 \lambda}})
  \sinh(\qh +i{\pi \over 4})} \over
 {\cosh(\qh -i{{j \pi} \over {4 \lambda}})
  \cosh(\qh +i{\pi \over 4}+i{{j \pi} \over {4 \lambda}})
  \sinh(\qh -i{\pi \over 4})}} \nonu
\times \prod_{k=1}^{j-1}
{{\sinh(\th+i{{k\pi}\over{2\lambda}})
  \cosh^2(\qh-i{\pi\over 4}-i{{k\pi}\over{4\lambda}})} \over
 {\sinh(\th-i{{k\pi}\over{2\lambda}})
  \cosh^2(\qh+i{\pi\over 4}+i{{k\pi}\over{4\lambda}})}} \ . \ \label{sG1}
\eea
The coupling parameters $(\eta,\vartheta)$ are contained in 
$R_1^j(\th)$, which can be written as
\be
R_1^j(\th)=S^j(\eta,\th)S^j(i\vartheta,\th) \ , \ \label{sG2}
\ee
where $S^j(x,\th)$ depends wether $j$ is even or odd. For
$j=2k$, $k=1,2,\ldots <{\lambda\over2}$, we have
\be
S^{2k}(x,\th)=\prod_{l=1}^k
{{\sinh(\th)-i\cos({x\over\lambda}-(l-{1\over2}){\pi\over\lambda})} \over
 {\sinh(\th)+i\cos({x\over\lambda}-(l-{1\over2}){\pi\over\lambda})}}
{{\sinh(\th)-i\cos({x\over\lambda}+(l-{1\over2}){\pi\over\lambda})} \over
 {\sinh(\th)+i\cos({x\over\lambda}+(l-{1\over2}){\pi\over\lambda})}} \ . \
\label{sG3}
\ee
For $j=2k-1$, $k=1,2,\ldots<{{\lambda+1}\over2}$, we have
\be
S^{2k-1}(x,\th)=
{{i \cos({x\over\lambda})-\sinh(\th)}\over
 {i \cos({x\over\lambda})+\sinh(\th)}}
\prod_{l=1}^{k-1}
{{\sinh(\th)-i\cos({x\over\lambda}-{{l\pi}\over\lambda})} \over
 {\sinh(\th)+i\cos({x\over\lambda}-{{l\pi}\over\lambda})}}
{{\sinh(\th)-i\cos({x\over\lambda}+{{l\pi}\over\lambda})} \over
 {\sinh(\th)+i\cos({x\over\lambda}+{{l\pi}\over\lambda})}} \ . \
\label{sG4}
\ee

\section{Reflection Matrices for the Supersymmetric FKM Models}

In the previous section we obtained the reflection matrices for the
FKM models. In order to write down the complete reflection matrix
for the supersymmetric generalizations we just have to ``attach" the
$R_{BF}(\theta)$ as given in (\ref{e3}), giving
\be
R^{[j]}(\th)= R^j_{FKM}(\th) \, Z^{(\pm)}_j(\th)
\left(\begin{array}{cc}
      \cosh({\th \over 2} \pm i {\pi \over 4}) & 0 \\
      0 & \cosh({\th \over 2} \mp i {\pi \over 4})
      \end{array}\right)\ \ , \ \label{xx}
\ee
where $R^j_{FKM}$ stands for either $R^j_{(1)}$ or
$R^j_{(2)}$ of (\ref{r1r2}). Before continuing 
we should point out that
the choice of $(1)$ or $(2)$ in the bosonic factor
is correlated with the choice of $(+)$ or $(-)$ in the
supersymmetric factor. We already mentioned that the
solution $R_{(1)}$ (which is the one with non-vanishing
boundary couplings) should be combined with 
$R_{BF}^{(-)}$, and it is then natural to associate 
$R_{(2)}$ with $R_{BF}^{(+)}$. We thus
propose the following complete reflection matrices
\be
R_{(1)}^{[j]} = R^j_{(1)} \, R_{BF}^{(-)} \ ,
\qquad
R_{(2)}^{[j]} = R^j_{(2)} \, R_{BF}^{(+)} \ .
\ee

Until now we have not specified the normalization
factors $Z_j^{(\pm)}(\theta)$ of the supersymmetric 
reflection matrices $R_{BF}^{(\pm)}$. Up to CDD factors, 
these are fixed by the equations (\ref{bdr11}) that express
unitarity and boundary crossing unitarity. We shall
now present explicit expressions for these factors.

The right hand side of the second equation of (\ref{bdr11})
depends on the mass $m_j$ of the reflecting particle
via the combination $\beta j$, where $\beta=1/(2n+1)$
for the susy FKM series, $\beta = 1/(2\lambda)$ for
susy sine-Gordon and $\beta \rightarrow 0$ for the
free theory.

Let us consider separately the two possibilities `$(-)$' 
and `$(+)$'. In the first case it is convenient to write 
$Z_j^{(-)}(\th)$ as
\be
Z_j^{(-)}(\th)={1 \over {\cosh({\th \over 2}-i {{\pi}\over4})}}
{\widetilde{Z}_j^{(-)}(\th)} \ , \
\ee
with $\widetilde{Z}_j^{(-)}(\th) \widetilde{Z}_j^{(-)}(-\th) = 1$.
The second equation of (\ref{bdr11}) then leads to
\be
{{\widetilde{Z}_j^{(-)}(\ipi-\th)}\over
{\widetilde{Z}_j^{(-)}(\ipi+\th)}}=
{\rm  exp}\left(i \int_{0}^{\infty}{dt \over t}
{{\sinh(\beta j t) \sinh((1-\beta j)t)} \over {\cosh^2({t \over 2}) 
\cosh(t)}} \sin({2 t \theta \over \pi})\right ) \ . \
\ee
By elementary methods we solve for ${\widetilde{Z}_j^{(-)}(\th)}$
and obtain
\be
Z_j^{(-)}(\th)={1 \over {\cosh({\th \over 2}-i {{\pi}\over4})}}
{\rm  exp}\left(-{i\over2} \int_{0}^{\infty}{dt \over t}
{{\sinh(\beta j t) \sinh((1-\beta j)t)} \over {\cosh^2({t \over 2}) 
\cosh^2(t)}} \sin({2 t \theta \over \pi})\right ) \ . \
\ee

In an analogous fashion we make the following
Ansatz for $Z_j^{(+)}(\theta)$
\be
Z_j^{(+)}(\th)={1 \over \cosh({\th \over 2}+ i {{\pi}\over4})}
\widetilde{Z}_j^{(+)}(\th) \ , \
\ee
with $\widetilde{Z}_j^{(+)}(\th) \widetilde{Z}_j^{(+)}(-\th) = 1$.
This leads to
\be
{{\widetilde{Z}_j^{(+)}(\ipi-\th)}\over{\widetilde{Z}_j^{(+)}(\ipi+\th)}}=
{\sinh(\th)-i\sin(\beta j \pi) \over \sinh(\th)+i\sin(\beta j \pi)}
\,
{\rm  exp}\left(i \int_{0}^{\infty}{dt \over t}
{{\sinh(\beta j t) \sinh((1-\beta j)t)} \over {\cosh^2({t \over 2}) 
\cosh(t)}} \sin({2 t \theta \over \pi})\right ) \ . \
\ee
We can rewrite the prefactor by using that for $\th \neq 0$ and real%
\footnote{This will be sufficient for the purpose
of boundary TBA calculations.}
\be
{{\sinh(\th)-i\sin(\alpha\pi)}\over{\sinh(\th)+i\sin(\alpha\pi)}}=
{\rm exp}\left(4i\int_0^\infty {{dt} \over t}
{{\cosh({\alpha\over2}t) \cosh({{1-\alpha}\over2}t)}\over{\cosh({t\over2})}} 
  \sin({{t\th}\over\pi}) \right) \ , \
\ee
and by the same elementary methods we obtain
\bea
Z_j^{(+)}(\th)={1 \over {\cosh({\th \over 2}+ i {{\pi}\over4})}} \;
&& {\rm exp}\left(-2i\int_0^{\infty}{{dt} \over t}
{{\cosh({1 \over 2} \beta j t) \cosh({1 \over 2}(1-\beta j)t)}
 \over {\cosh^2({t\over2})}} \sin({{t\th}\over\pi})
\right) \nonu
\times \, && {\rm  exp}\left(-{i\over2} \int_{0}^{\infty}{dt \over t}
{{\sinh(\beta j t) \sinh((1-\beta j)t)} \over {\cosh^2({t \over 2}) 
\cosh^2(t)}} \sin({2 t \theta \over \pi})\right )\ . \
\eea

Note that in the free limit, $\beta \rightarrow 0$, these
$Z$-factors reproduce the ones that we used in (\ref{ee}).

\section{Reflection matrices for Bound States of Susy sine-Gordon}

In ref.~\cite{IOZ}, it was found that there are two special
choices for boundary conditions on the {\em classical}\ susy 
sine-Gordon theory such that the resulting theory is both integrable 
and supersymmetric. Expressed in the variables of the 
action (\ref{ssG1}), these conditions are
\be
{\cal{BC}}^{\pm}: \qquad 
\partial_x \phi \pm {2m \over \beta_{ssG}} 
  \sin ({\beta_{ssG} \, \phi \over 2}) = 0 \ ,
\qquad \psi \mp \bar\psi = 0
\ee
at $x=0$. In the notation of \cite{GZ}, the bosonic part of
these conditions corresponds to $M= M_{\pm} = 
\pm {4m \over \beta^2}$, $\phi_0=0$. We shall refer to these 
boundary conditions as ${\cal{BC}}^{\pm}$.

We conjecture that at those two special points,
the reflection matrix for the bound state multiplets
$(b_j,f_j)$, $j=1,2,\ldots < \lambda$ will be of the form 
$R_B \otimes R_{BF}$, where $R_B$ is a reflection
matrix for the breathers of the sine-Gordon theory,
see (\ref{sG1}) -- (\ref{sG4}) and $R_{BF}$ is the
supersymmetric reflection matrix of section III.

Two questions arise: what are the values of $\eta$
and $\vartheta$ for these two special points, and how can
we match the choice of sign in the boundary conditions
${\cal{BC}}^{\pm}$ with the choice of sign in the
reflection matrix $R_{BF}^{(\pm)}$?

Part of the answer to the first question can be found
in \cite{GZ}, where it is shown that the $\phi_0=0$ boundary 
potentials correspond to a condition $\xi=0$ that is satisfied
when $\vartheta=0$. If we now look at the Ghoshal 
reflection matrix $R^1_{sG}(\theta | \eta,\vartheta=0)$ 
(for the first breather, $j=1$) for generic $\eta$, 
we see that it has zeros at $\theta = i {\pi \over 2}$, 
$i \pi {2\lambda-1 \over 2\lambda}$, 
$i \pi {1-\lambda \over 2 \lambda}$. We can then look 
for special values of $\eta$ that are such that one of 
these zero's gets canceled. This happens when $\eta=\lambda \pi$, 
${\pi \over 2}(\lambda+1)$, ${\pi \over 2}$, resp. 
The point $\eta = {\pi \over 2}(\lambda + 1)$ has been 
identified with the value $M=0$ (free boundary conditions 
in the bosonic sector), see \cite{GZ}.

At the special values $\lambda={2n+1 \over 2}$ the minimal
choices $\eta = \lambda \pi$ and $\eta = {\pi \over 2}$
precisely give the two minimal reflection matrices
$R_{(1)}$ and $R_{(2)}$ for the $n$-cpt FKM minimal models 
as given in our section IV. 

Combining these observations, it is natural to conjecture that 
the two points that correspond to an integrable, supersymmetric 
boundary potential are precisely the two points $\eta=\lambda\pi$, 
${\pi \over 2}$ where the reflection matrix is `minimal' with 
non-zero $M$.

What remains to be specified is which of the two factors 
$R_{BF}^{(\pm)}(\theta)$ should be used in the two cases 
at hand, and also what the correspondence with the boundary 
conditions ${\cal{BC}}^{\pm}$ is. The minimal solution with 
$\eta=\pi \lambda$, called $R_{(1)}$ in the context of the 
FKM minimal models, has non-vanishing boundary couplings and should 
thus be combined with $R_{BF}^{(-)}$ (see section III).
This leaves $R_{BF}^{(+)}$ to be combined with the
solution at $\eta= {\pi \over 2}$
 
Let us now recall that the ratio $R_f/R_b$ of the amplitudes
in the factor $R_{BF}^{(\pm)}$ precisely corresponds
to the reflection amplitude for a Majorana fermion,
with free boundary condition, $\psi-\bar\psi = 0$,  
in the case of $R_{BF}^{(+)}$ and fixed boundary condition,
$\psi+\bar\psi = 0$, in the case of $R_{BF}^{(-)}$
(see section III). The fermionic components of the 
boundary conditions ${\cal{BC}}^{\pm}$ are precisely of this same form,
and this suggests that we link ${\cal{BC}}^+$ to $R_{BF}^{(+)}$
and ${\cal{BC}}^-$ to $R_{BF}^{(-)}$. This identification gets
further confirmed by the correspondence between the
our supercharges $Q^{(\pm)}$ and the susy transformations given
in \cite{IOZ}. We therefore expect that the 
complete result for the reflection matrices of the breather 
multiplets of susy sine-Gordon are
\bea
&& R_{sG}(\theta | \eta={\pi \over 2}, \vartheta=0) \, R^{(+)}_{BF}(\theta)
 \qquad {\rm for} \qquad {{\cal{BC}}^+}
\nonu
&& R_{sG}(\theta | \eta=\lambda \pi, \vartheta=0) \, R^{(-)}_{BF}(\theta)
 \qquad {\rm for} \qquad {{\cal{BC}}^-} \ .
\eea
At the special values $\lambda = {2n+1 \over 2}$ these reflection
matrices reduce to the two supersymmetric reflection matrices 
that we propose for the susy FKM models. 

\section{Conclusions}

We have proposed exact reflections matrices for a number of
integrable $N=1$ supersymmetric theories. The examples studied 
are the supersymmetric sine-Gordon theory and a series
of minimal models (called susy FKM) that arise as perturbations of 
superconformal field theories. These theories all have exact (bulk) 
$S$-matrices that enjoy the factorization as written in (\ref{d}). 

The main result of this paper is the structure of the reflection 
matrix (\ref{e3}), which was obtained directly from the following
two assumptions: (i) both integrability and supersymmetry
are preserved after the introduction of the boundary and 
(ii) the boundary has no structure, that is, it does not allow 
scattering processes that violate fermion number.

In order to write down the complete reflection matrix for the models 
studied in this paper we started from the reflection matrices for 
their bosonic reductions. In the case of sine-Gordon model, these were
given by Ghoshal in \cite{Gh}. For the susy FKM models such reflection
matrices are for the first time given in this paper. We derived them 
by exploiting a connection with the $A_{2n}^{(2)}$ twisted affine Toda 
theories.

In \cite{IOZ} Inami, Odake and Zhang proposed two possible boundary 
conditions that preserve supersymmetry and integrability upon the 
introduction of a boundary in the susy sine-Gordon model. Their result  
was based on a classical analysis of the constraints imposed by the first 
non-trivial conserved charge. We conjectured a connection between the 
reflection matrices found in this paper and the boundary conditions 
proposed in \cite{IOZ}.

There are a number of issues that deserve further attention. 
It would be interesting to understand better the origin of the
two distinct reflection matrices $R_{(1)}$ and $R_{(2)}$ for
the susy FKM models, for example by first studying boundary
fields in the unperturbed superconformal field theory.
Another problem that deserves consideration would be to perform a 
TBA analysis, generalizing the methods used in \cite{MS,Ahn1} from 
the bulk system to the boundary system, and to investigate the 
connections with $N=2$ theories as was done in \cite{MS} for the 
bulk theory. One may also consider more general (possibly
non-supersymmetric) solutions to the boundary Yang-Baxter
equations for the $S$-matrices studied in this paper. 

\section{Acknowledgements}
KS thanks T.~Inami for hospitality at the Yukawa Institute and
for explaining the results of \cite{IOZ}. KS thanks Princeton
University for hospitality and support. MM thanks R. K\"oberle and 
A. Fring for useful discussions and the University of Amsterdam for 
hospitality. The research of KS is supported in part by the foundation 
FOM (the Netherlands), while MM was partially supported by CNPq (Brazil).

\end{document}